\documentclass[letterpaper]{article}
\usepackage{uai2020}
\usepackage{todonotes}
\usepackage{amsmath, mathtools}
\usepackage{centernot}
\usepackage{amssymb}
\usepackage{graphicx,subfigure}
\usepackage{amsthm}
\usepackage{tabu}
\usepackage{paralist}
\usepackage{physics}
\usepackage{caption}
\usepackage{accents}
\usepackage{natbib}

\newtheoremstyle{break}
  {\topsep}{\topsep}%
  {\itshape}{}%
  {\bfseries}{}%
  {\newline}{}%
\theoremstyle{break}

\newtheorem{theorem}{Theorem}

\newcounter{example}[section]
\newenvironment{example}[1][]{\refstepcounter{example}\par\medskip
   \noindent \textbf{Example~\theexample. #1} \rmfamily}{\medskip}

\newcommand{\bs}{\boldsymbol}
\newcommand{\given}{\:\vert\:}
\newcommand{\mc}{\mathcal}
\newcommand{\ci}{\mathrel{\perp\mspace{-10mu}\perp}}
\newcommand\numberthis{\addtocounter{equation}{1}\tag{\theequation}}
\newcommand{\nci}{\centernot{\ci}}
\DeclareMathOperator{\EX}{\mathbb{E}}
\newcommand{\Cov}{\mathrm{Cov}}
\newcommand{\mtx}[1]{\begin{bmatrix} #1 \end{bmatrix}}
\newcommand{\rvector}[1]{%
  \begingroup\def\arraystretch{0}\begin{bmatrix}
  #1
  \end{bmatrix}\endgroup}
  
\usepackage[margin=1in]{geometry}
\usepackage{hyperref}
\hypersetup{colorlinks,linkcolor={blue},citecolor={blue},urlcolor={red}}  

\usepackage{times}

\title{Learning Joint Nonlinear Effects from Single-variable Interventions in the Presence of Hidden Confounders}

\author{ {\bf Sorawit Saengkyongam } \\
University College London \\
\And
{\bf Ricardo Silva}  \\
University College London \\
The Alan Turing Institute
}

\begin{document}

\maketitle

\begin{abstract}
We propose an approach to estimate the effect of multiple simultaneous interventions in the presence of hidden confounders. To overcome the problem of hidden confounding, we consider the setting where we have access to not only the observational data but also sets of \emph{single-variable} interventions in which each of the treatment variables is intervened on separately. We prove identifiability under the assumption that the data is generated from a nonlinear continuous structural causal model with additive Gaussian noise. In addition, we propose a simple parameter estimation method by pooling all the data from different regimes and jointly maximizing the combined likelihood. We also conduct comprehensive experiments to verify the identifiability result as well as to compare the performance of our approach against a baseline on both synthetic and real-world data.
\end{abstract}

\section{MOTIVATION \& CONTRIBUTION}
\label{introduction}

To adopt data-driven approaches in many scientific or business domains, one often has to deal with causal inference problems where the causal effects of different treatments are of interest. Estimating causal effects purely from observational data, however, often requires strong untestable assumptions. One such major assumption is an absence of \emph{hidden confounders} (unobserved common causes) which is often violated in practice. For instance, consider the problem of inferring the effect of medical treatments from past observational data. It is difficult to account for all confounders since we often do not know precisely how the doctors prescribed the treatments in the first place.

A gold standard method to overcome the issue of hidden confounding is to conduct a controlled experiment in which the variables of interest are directly intervened by the examiner, and thus all of the confounders are fully observed by design or eliminated altogether (in the case of randomized experiments). In the previous example, one could conduct a randomized controlled trial where the treatments are randomly assigned to the patients. By doing so, all of the confounders are eliminated and treatment effects can be estimated with standard statistical inference methods.

In many real-world scenarios, however, we may not be able to obtain experimental data due to feasibility issues such as budget, time, and ethical constraints. These feasibility constraints are especially more prominent when we consider the effect of joint interventions. One such example is multiple gene knockout experiments where we aim to estimate the effect of simultaneous gene knockouts on a phenotype of interest. The number of possible combinations of simultaneous knockouts increases exponentially with the number of genes. Conducting such a large number of knockout experiments are, therefore, not practically feasible. The problem is even more prominent when there are more than two treatment levels per variable. In this case, using causal models to predict joint interventional effects from observational data is an attractive option. However, as mentioned before, existing methods often assume the absence of hidden confounders for some pairs of variables, or return uninformative results if such a condition is not satisfied. To remove this strong assumption, we consider a setting where, in addition to the observational data, we have access to sets of \emph{single-variable} interventional data in which each treatment variable is intervened on separately. We believe this is a practically reasonable setting since the number of interventional experiments needed only grows linearly with the number of treatment variables. \\

\begin{example}
To illustrate the setting of our problem, we consider the following hypothetical example with two treatment variables. A biologist seeks to predict a phenotypic
response ($Y$) to simultaneous perturbations of genes $X_1$ and $X_2$, where all measurements are continuous. Because of some practical constraints, the biologist can only obtain data generated from single-gene perturbation experiments in addition to the observational data. In other words, the biologist is given observational data where $Y, X_1$, $X_2$ are distributed according to the natural regime, and two sets of interventional data where $X_1$ and $X_2$, are being perturbed separately, at different levels. The goal is to predict the effect of joint interventions at combined levels of genes $X_1$ and $X_2$ on the phenotype $Y$. Figure \ref{fig:2vars} depicts a graphical representation of this simplified example.
\end{example}

\textbf{Contribution:} In this work, we consider a problem of learning joint interventional effects in the presence of hidden confounders from a combination of observational data and \emph{single-variable} interventions. We first show that, without any restrictions on the underlying structural equations, the effect of joint interventions is not identifiable. However, by introducing some reasonably weak assumptions, namely additive noise models, the effect is then identifiable. Furthermore, we introduce an inference algorithm to learn the parameters of the proposed model by pooling data across different regimes and jointly maximizing the combined likelihood. Our result provides the means to dramatically reduce the number of interventional experiments needed to estimate the effect of joint interventions compared to a blackbox model.


\section{BACKGROUND \& RELATED WORK}
\label{sec:bg}
\subsection{BACKGROUND}
\textbf{Structural Causal Models:}
Our work relies on the structural causal model (SCM) framework described by \citet{PE09}. SCMs allow us to encode causal assumptions by describing the data generating process not only in terms of probability distributions, but also its causal mechanisms through a set of structural equations. Specifically, an SCM $\mc{M}$ is a 4-tuple $\langle \bs{X},\bs{U}, \bs{f} ,P_{\bs{U}} \rangle$, where
\begin{compactenum}
    \item $\bs{X}$ represents endogenous variables.
    \item $\bs{U}$ represents exogenous (``noise'') variables.
    \item $\bs{f}$ is a set of structural equations, each of which describing how the corresponding endogenous variable causally depends on other variables, i.e. $X_i = f_i(\textbf{PA}_{X_i}, \bs{U}_i)$ (the set $\textbf{PA}_{X_i} \in \bs{X}\backslash{X_i}$ are called \emph{parents} or \emph{direct causes} of $X_i$, and $\bs{U}_i \in \bs{U}$ are the corresponding noise variables).
    \item $P_{\bs{U}}$ is a joint probability distribution over the noise variables $\bs{U}$.
\end{compactenum}
An SCM $\mc{M}$ entails a joint distribution over the endogenous variables denoted by $P^{\mc{M}}_{\bs{X}}$. If $\mc{M}$ is unaffected by any intervention, we refer to the entailed distribution as the \emph{observational distribution}. An SCM also induces a unique causal graph $G$. Each node in $G$ represents one of the endogenous variables, while edges denote the causal mechanisms defined by the structural equations $\bs{f}$. Specifically, the directed edge is drawn from $X_j$ to $X_i$ if $X_j \in \textbf{PA}_{X_i}$. Furthermore, if exogenous variables are not jointly independent, a bi-directed edge will be drawn between the corresponding pairs of endogenous variables. In other words, a bi-directed edge is drawn between $X_i$ and $X_j$ if $\bs{U}_i \nci \bs{U}_j$. These dependencies between exogenous variables indicate the presence of \emph{hidden confounders}.

\textbf{Perfect Intervention:}
A crucial property of SCMs is the invariance of structural equations, that is, each structural equation is unaffected by any changes made on other structural equations. This property allows us to define a mathematical operator for performing perfect intervention called the $``do(\cdot)"$ operator. Applying $``do(X_i = x)"$ operator on an SCM $\mc{M}$ results in a post-intervened SCM $\mc{M}_{do(X_i=x)}$ in which the original equation for $X_i$ is replaced by the constant $x$. From $\mc{M}_{do(X_i)}$, we can obtain the entailed distribution $P^{\mc{M}_{do(X_i)}}_{\bs{X}}$. This is referred
to as an \emph{interventional distribution}.

\begin{figure}[t!]
    \centering
    \subfigure[]
    {
        \includegraphics[scale=.305]{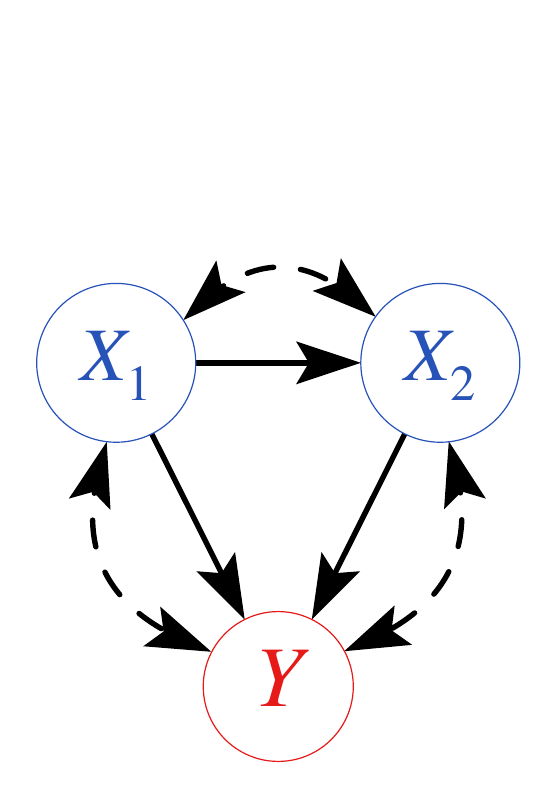}
    }
    \subfigure[]
    {
        \includegraphics[scale=.305]{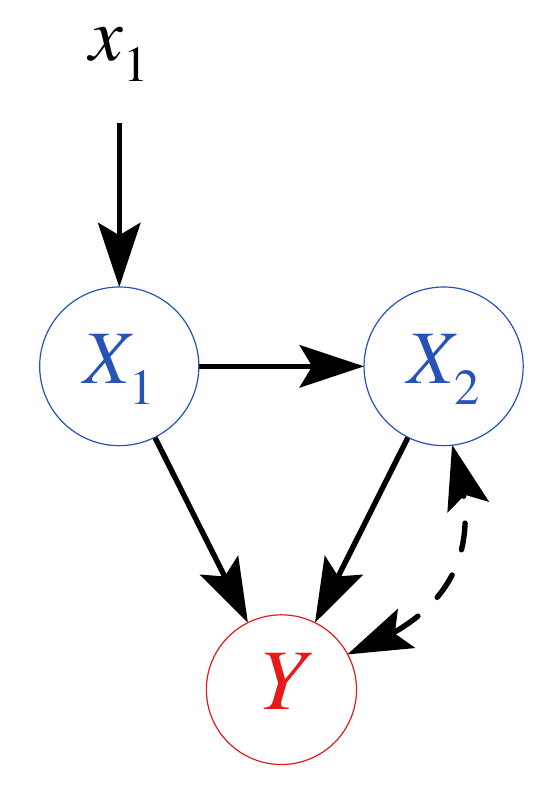}
    }
    \subfigure[]
    {
        \includegraphics[scale=.305]{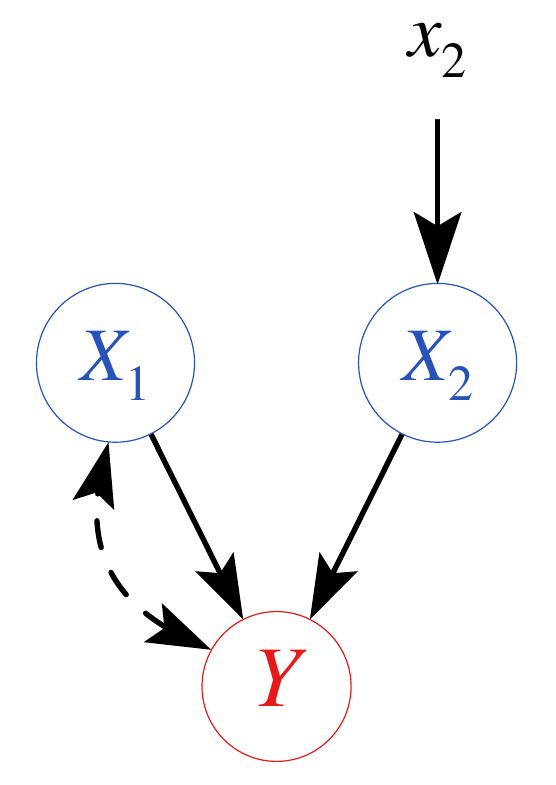}
    }
    \subfigure[]
    {
        \includegraphics[scale=.305]{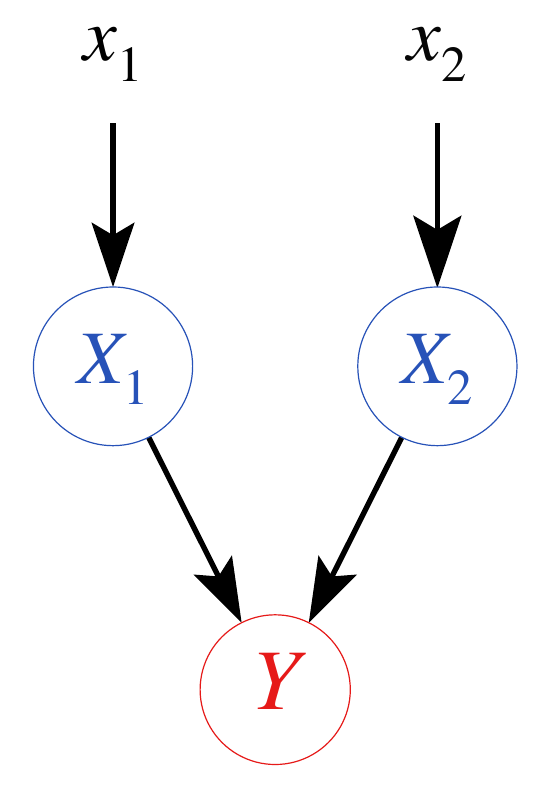}
    }
    \caption
    {An example with two treatment variables. (a) the causal graph associated with the unintervened SCM $\mc{M}$. (b, c) the causal graphs associated with the singly-intervened SCMs $\mc{M}_{do(X_1)}$, $\mc{M}_{do(X_2)}$. (d) the causal graph associated with the jointly-intervened SCM $\mc{M}_{do(X_1, X_2)}$. We are given data from (a), (b), (c) and aim to make predictions in (d).}
    \label{fig:2vars}
\end{figure}

\subsection{RELATED WORK}
\textbf{Identification of nonparametric SCMs:} The identification of causal effects under nonparametric structural causal models has been extensively studied over the past decades. In the nonparametric case, the problem of identification asks whether the causal queries can be uniquely computed from available data and a given causal graph regardless of the underlying structural equations and the distribution of the noise variables. The early work focused on identifying causal effects purely from observational data \citep{PE95, TIPE02, HUEA08}. Later on, \citet{ELPE12} extended the notion of identifiability to consider not only observational data but also interventional data called z-identifiability. Recently, more general results were introduced by \citet{LEEA19} in which a necessary and sufficient graphical condition for identifying causal effects from arbitrary combinations of observational and interventional distributions was provided. 

Despite the great progress in nonparametric identification over the past decade, causal effects in many real-world settings are not identifiable under this framework due to its reliance on independence constraints only. In particular, we show in Section \ref{sec:unindent} that the effect of joint interventions is not nonparametrically identifiable under our problem setting. 

\textbf{Additive Noise Models:} By imposing some restrictions on the structural equations, one could obtain more powerful identifiability results. One of the most commonly used restrictions are the additive noise models (ANMs) \citep{HOEA09} which limit the form of the structural equations to be additive with respect to the noise variables. Based on ANMs, \citet{JA09} proposed a method for inferring a latent confounder between two observed random variables which is otherwise not possible without additional assumptions. Concerning the task of learning causal structure, ANMs also permit us to recover the underlying causal graph from observational data beyond the Markov equivalence class \citep{ZHAA09,JOEA11,PEEA13, PEEA14}. Inspired by the previous successes, our work relies on this model to provide identifiability of joint interventional effects.



\textbf{Effect Estimation in SCMs:} 
Related to our problem, \citet{NAEA17} developed a method to estimate the effect of joint interventions from observational data when the causal structure is not known, \textsf{joint-IDA}, which is an extension of the \textsf{IDA} algorithm of \citet{MAEA09}. In \textsf{IDA}, the assumed causal structure is obtained via a causal discovery algorithm that returns all graphs in the Markov equivalence class. The algorithm exploits properties of a linear SCM with Gaussian noise to avoid the na\"{i}ve approach of enumerating all graphs in the Markov equivalence class. Hidden confounders are allowed in the extension by \citet{MAL16}, where the Markov equivalence class is less informative and a reasonable degree of sparsity (in the observable marginal distribution) is to be expected if nontrivial solution sets are to be returned. Another closely related work is the work of \citet{HYEA12}. They considered the problem of learning the structure and parameters of linear cyclic SCMs, where they make use of interventional data to remove the hidden confounding assumption. However, their approach relies on the linearity assumption, which is a particularly strong restriction.

The goal of our work is broadly similar to that by \citet{NAEA17} and \citet{HYEA12}. However, we drop the linearity assumption used in both of the works. Besides, no sparsity is required for identification. The price to be paid is our requirement for single-variable interventions to be observed, which we believe is a reasonable setting in important real-world applications. This complements the literature on methods for sparse Markov equivalence classes with purely observational data. 

\section{PROBLEM SETTING}
\label{sec:problem_def}
We now define the setting we consider in our work.

Let $\bs{X} = \{X_1, \dots, X_K\}$ be a set of treatment variables and $Y$ be an outcome variable. Denote $\bs{X}^k := \{X_i\}_{i=1}^k$. \\ $\bs{X}$ and $Y$ are assumed to be generated from an SCM $\mc{M},$
\begin{align*}
    Y   &= f_Y(\bs{X}^K, U_Y) \\
    X_k &= f_k(\bs{X}^{k-1}, U_k), \, \text{for}\, k = 1, \dots, K, \numberthis \label{eq:3.2}
\end{align*}
where $U_1, \dots, U_Y$ are the noise variables (note that we do not assume the noise variables to be mutually independent).

We are given i.i.d. samples from observational regime $D_{obs} \sim P^{\mc{M}}_{(\bs{X}, Y)}$. In addition, we are also given a set of i.i.d. samples from single-variable interventional regimes where we intervene on $X_1, \dots, X_K$ separately, i.e. $\bs{D}_{int} =  \{D^k_{int} \sim P^{\mc{M}_{do(X_k)}}_{(\bs{X}, Y)} \}_{k=1}^K$.

The causal query of interest is the effect of joint interventions in terms of the conditional expectation $
Q(\mc{M}) = \EX[Y \given do(X_1, \dots, X_K)]$.

Figure \ref{fig:2vars} illustrates an example with two treatment variables.

\section{UNIDENTIFIABILITY UNDER UNCONSTRAINED SCMs}
\label{sec:unindent}
We will now show that, without additional assumptions on the form of structural equations, the joint interventional effects are not identifiable. To illustrate the unidentifiability, we consider the case where there are two treatment variables. We will show that there exists a pair of SCMs $\ddot{\mc{M}}$, $\tilde{\mc{M}}$ such that they entail identical observational distribution ($P^{\ddot{\mc{M}}}= P^{\tilde{\mc{M}}}$) as well as identical single-variable interventional distributions ($P^{\ddot{\mc{M}}_{do(X_1)}} = P^{\tilde{\mc{M}}_{do(X_1)}}$ and $P^{\ddot{\mc{M}}_{do(X_2)}} = P^{\tilde{\mc{M}}_{do(X_2)}}$), but induce different joint interventional distributions ($P^{\ddot{\mc{M}}_{do(X_1, X_2)}}~\neq~P^{\tilde{\mc{M}}_{do(X_1, X_2)}}$).

Define SCMs $\ddot{\mc{M}}$, $\tilde{\mc{M}}$ as follows,
\begin{table}[h]
    \centering
\begin{tabular}{c|c}
$\ddot{\mc{M}}$ & $\tilde{\mc{M}}$ \\ \hline
{$\!\begin{aligned}
    Y   &= X_1 \land X_2 \land U_Y \\
    X_2 &= X_1 \land U_2 \\
    X_1 &= U_1
\end{aligned}$} & {$\!\begin{aligned}
    Y   &= X_2 \land U_Y \\
    X_2 &= X_1 \land U_2 \\
    X_1 &= U_1
\end{aligned}$}
\end{tabular}
\vskip -0.1in
\end{table}

where $U_Y =  U_2 = U_1 \sim \text{Bernoulli}(p).$

\begin{table}[t!]
\centering
\captionsetup{font=footnotesize}
\caption{Observational Distribution: $P^{\ddot{\mc{M}}}, P^{\tilde{\mc{M}}}$}
\vskip -0.1in
\scalebox{1.0}{
\begin{tabular}{c|cc}
$P(X_1, X_2, Y)$ & $Y=0$ & $Y=1$ \\ \hline
$X_1, X_2 = 0,0$ & $1-p$ & 0   \\
$X_1, X_2 = 1,0$ & 0   & 0   \\ 
$X_1, X_2 = 0,1$ & 0   & 0   \\
$X_1, X_2 = 1,1$ & 0   & $p$ \\ 
\end{tabular}
}
\label{table:obs}
\end{table}

\begin{table}[t!]
\centering
\captionsetup{font=footnotesize}
\caption{Interventional Distribution: $P^{\ddot{\mc{M}}_{do(X_1)}}, P^{\tilde{\mc{M}}_{do(X_1)}}$}
\vskip -0.1in
{\tabulinesep=0.7mm
\scalebox{1.0}{
    \begin{tabu}{c|c}
    $do(X_1=0)$     &  \begin{tabular}{c|cc}
    $P(X_2, Y)$ & $Y=0$ & $Y=1$ \\ \hline
    $X_2 = 0$ & 1 & 0   \\
    $X_2 = 1$ & 0   & 0   \\ 
    \end{tabular} \\ \hline 
    $do(X_1=1)$ & \begin{tabular}{c|cc}
    $P(X_2, Y)$ & $Y=0$ & $Y=1$ \\ \hline
    $X_2 = 0$ & $1-p$ & 0   \\
    $X_2 = 1$ & 0   & $p$   \\ 
    \end{tabular}
    
    \end{tabu}
    }
}
\label{table:intx1}
\end{table}

\begin{table}[t!]
\centering
\captionsetup{font=footnotesize}
\caption{Interventional Distribution: $P^{\ddot{\mc{M}}_{do(X_2)}}, P^{\tilde{\mc{M}}_{do(X_2)}}$}
\vskip -0.1in
{\tabulinesep=0.7mm
\scalebox{1.0}{
    \begin{tabu}{c|c}
    $do(X_2=0)$     &  \begin{tabular}{c|cc}
    $P(X_1, Y)$ & $Y=0$ & $Y=1$ \\ \hline
    $X_1 = 0$ & $1-p$ & 0   \\
    $X_1 = 1$ & $p$   & 0   \\ 
    \end{tabular} \\ \hline 
    $do(X_2=1)$ & \begin{tabular}{c|cc}
    $P(X_1, Y)$ & $Y=0$ & $Y=1$ \\ \hline
    $X_1 = 0$ & $1-p$ & 0   \\
    $X_1 = 1$ & 0   & $p$   \\ 
    \end{tabular}
    
    \end{tabu}
    }
}
\label{table:intx2}
\end{table}

Tables \ref{table:obs}, \ref{table:intx1}, \ref{table:intx2} show the joint probabilities in the observational and single-variable interventional regimes. The two given SCMs yield identical observational distributions and identical single-variable interventional distributions. However, they yield different joint interventional distributions,
\begin{align*}
    P^{\ddot{\mc{M}}_{do(X_1=0, X_2=1)}}(y) =\left\{
    \begin{array}{@{}ll@{}}
    1, & \text{if}\ y = 0 \\
    0, & \text{otherwise}
    \end{array}\right. \\
    \neq \, P^{\tilde{\mc{M}}_{do(X_1=0, X_2=1)}}(y) =\left\{
    \begin{array}{@{}ll@{}}
    p, & \text{if}\ y = 1 \\
    1-p, & \text{if}\ y = 0 \\
    0, & \text{otherwise}
    \end{array}\right.
\end{align*}
Hence, the join interventional effects are non-identifiable under unconstrained SCMs.

\section{PROPOSED APPROACH}
\label{sec:approach}
The result in the previous section motivates us to consider a restricted class of causal models, particularly additive noise models. In additive noise models, we impose an additional assumption on the structural equations, that is, the noise variables (exogenous variables) affect the observables (endogenous variables) in an additive way. Considering our problem definition, we essentially replace \eqref{eq:3.2} with
\begin{align*}
    Y   &= f_Y(\bs{X}^K) + U_Y \\
    X_k &= f_k(\bs{X}^{k-1}) + U_k, \, \text{for}\, k = 1, \dots, K \label{eq:3.3} \numberthis{}
\end{align*}
while the rest remain the same. Furthermore, the noise distribution is assumed to be a zero-mean Gaussian: $(U_1, \dots, U_Y) \sim \mathcal{N}(\bs{0}, \Sigma)$, where $\Sigma$ is an arbitrary covariance matrix of size $(K+1) \times (K+1)$.

\subsection{MAIN ASSUMPTIONS} Our approach relies on the following main assumptions.
\begin{enumerate}
    \item (Additive noise) the underlying model is assumed to be an additive noise model.
    \item (Gaussian noise) the noise distribution is assumed to be a zero-mean Gaussian with arbitrary covariance matrix.
    \item (Known causal structure) the underlying causal structure is assumed to be given.
    \item (Acyclicity) the underlying causal structure can be represented by an acyclic directed mixed graph \citep[ADMG,][]{richardson:03}.
\end{enumerate}
Assumptions 1 and 2 play a major role in providing the identifiability of joint interventional effects. The main implication of the additive Gaussian noise assumption is that we can only consider continuous variables. Nonetheless, we believe that the Gaussianity assumption can be relaxed to some extent which may shed some light on how to handle discrete variables. We leave this for future work. Furthermore, we note that we assume Assumption 3 for clarity of exposition where we focus on the identifiability and estimation of the joint effects. However, it is conceptually straightforward to obtain an estimated causal structure in our setting as we have access to the single-variable interventions for all treatment variables \citep{EBEA06}. 

\subsection{MODEL IDENTIFICATION}
Our main theoretical result is the identification of the effect of joint interventions on all control variables $\bs X$ from a combination of observational data $D_{obs}$ and single-variable interventional data $\bs{D}_{int}$.

\begin{theorem}[Identifiability of joint nonlinear effects under additive noise models]
Let $\mc{M}^K = \langle \{\bs{X}, Y\}, \bs{U}, \bs{f}, P_{\bs{U}} \rangle$ be an additive noise SCM with $K$ treatment variables,
\begin{align*}
    Y   &= f_Y(\bs{X}^K) + U_Y \\
    X_k &= f_k(\bs{X}^{k-1}) + U_k, \, \text{for}\, k = 1, \dots, K
\end{align*}
with $P_{\bs{U}} \sim \mc{N}(\bs{0}, \Sigma)$, where $\Sigma$ is an arbitrary covariance matrix and $\bs{X}^k := \{X_i\}_{i=1}^k$. The causal query $Q(\mc{M}^K) = \mathbb E[Y \given do(X_1, \dots, X_K)]$ is identifiable from a combination of the observational distribution $P^{\mc{M}^K}_{(\bs{X}, Y)}$ and the set of single-variable interventional distributions $\{P^{\mc{M}^K_{do(X_i)}}_{(\bs{X}, Y)} \}_{i=1}^K$, for any integer $K \geq 2$.
\end{theorem}

We prove the theorem by induction. In the base case, we show that $\EX[Y \given do(X_1, X_2)]$ is identifiable. In the inductive step, given that $\EX[Y \given do(X_1, \dots, X_{K})]$ is identifiable, we show that $\EX[Y \given do(X_1, \dots, X_{K+1})]$ is also identifiable. A sketch of the proof for the base case is provided below. Please refer to the supplementary material for the full proof.
\begin{proof}[Proof sketch for the base case]
The query of interest is
\begin{equation*}
    Q(\mc{M}^2) = \EX[Y \given do(X_1 = x_1, X_2 = x_2)] = f_Y(x_1, x_2).
\end{equation*}
Due to unobserved confounders, the above query is not identifiable solely from the observational distribution $P^{\mc{M}^2}_{(\bs{X}, Y)}$,
\begin{align*}
    \EX[Y \given X_1 = x_1, X_2 = x_2]& \\
    = f_Y(x_1, x_2) + &\EX[U_Y \given X_1 = x_1, X_2 = x_2].
\end{align*}
However, if we are able to identify $\EX[U_Y \given X_1 = x_1, X_2 = x_2]$, we would then be able to identify our query of interest $f_Y(x_1, x_2)$. We then need to show that the expected noise $\EX[U_Y \given X_1 = x_1, X_2 = x_2]$ can be uniquely computed from a combination of $P^{\mc{M}^2}_{(\bs{X}, Y)}$, $P^{\mc{M}^2_{do(X_1)}}_{(\bs{X}, Y)}$ and $P^{\mc{M}^2_{do(X_2)}}_{(\bs{X}, Y)}$.

From the additive Gaussian noise assumptions, we have
\begin{equation}
    \EX[U_Y \given X_1 = x_1, X_2 = x_2] = \Sigma_{u_y}\Sigma_{u_x}^{-1}\bs{u}_x, \label{eq:2var_euy}
\end{equation}
with $\bs{u}_x = \rvector{x_1 & x_2 - f_2(x_1)}^\top$, $\Sigma_{u_y} = \rvector{\sigma_{Y1} & \sigma_{Y2}}$ and $\Sigma_{u_x} = \mtx{\sigma_{11} & \sigma_{12} \\
\sigma_{21} & \sigma_{22}}$,
where we define $\sigma_{ij} := \Cov(U_i, U_j)$.

From Equation \eqref{eq:2var_euy}, the quantities that we need to show the identifiability are $f_2$, $\Sigma_{u_x}$ and $\Sigma_{u_y}$. 

\subsubsection*{Identifying $f_2$ and $\Sigma_{u_x}$}
$f_2$ can be trivially obtained from $P^{\mc{M}^2_{do(X_1)}}_{(\bs{X}, Y)}$,
\begin{equation*}
    \EX[X_2 \given do(X_1 = x_1)] = f_2(x_1).
\end{equation*}
Since $f_2$ is identified, we can then identify the joint distribution $p(U_1, U_2)$ from $P^{\mc{M}^2}_{(\bs{X}, Y)}$,
\begin{align*}
    U_2 &= X_2 - f_2(X_1) \\
    U_1 &= X_1
\end{align*}
And thus, the covariance matrix $\Sigma_{u_x}$ is identifiable.

\subsubsection*{Identifying $\sigma_{Y1}$ and $\sigma_{Y2}$}
The idea is to identify these covariances by contrasting the expected treatment responses in the different interventional regimes. We give the sketch for the identification of $\sigma_{Y1}$ below ($\sigma_{Y2}$ can be obtained in a similar fashion).

From the regime $P^{\mc{M}^2_{do(X_1)}}_{(\bs{X}, Y)}$, we have
\begin{equation}
    \EX[Y \given do(X_1 = x_1)]
    = \EX[f_y(x_1, f_2(x_1)+ U_2)] \label{eq:idn-proof-1}
\end{equation}
From the regime $P^{\mc{M}^2_{do(X_2)}}_{(\bs{X}, Y)}$, we have
\begin{equation*}
\begin{split}
    \EX[Y \given X_1 = x_1, do(X_2 = x_2)]& \\
    = f_y(x_1, x_2) + &\EX[U_Y \given X_1 = x_1]
\end{split}
\end{equation*}
We can hypothetically choose $x_2 = f_2(x_1) + u_2$ and treat the above solely as a mathematical expression that can take a random variable as an input, in this case $U_2$:
\begin{equation*}
\begin{split}
    \EX[Y \given X_1 = x_1, do(X_2 = f_2(x_1) + U_2)].
\end{split}
\end{equation*}
We then take its expectation with respect to the identifiable marginal $p(U_2)$. The left-hand side of the equation below is also identifiable since $f_2$ is and we observe all single-variable interventions.
\begin{equation}
\begin{split}
    \EX_{p(U_2)}[\EX[Y \given X_1 = x_1, do(X_2 = f_2(x_1) + U_2)]]& \\
    = \EX[f_y(x_1, f_2(x_1) + U_2)] + \EX[U_Y \given X_1 = x_1]. \label{eq:idn-proof-2}
\end{split}
\end{equation}
Subtracting \eqref{eq:idn-proof-1} from \eqref{eq:idn-proof-2}, we get
\begin{equation*}
    \eqref{eq:idn-proof-2}-\eqref{eq:idn-proof-1} =
    \EX[U_Y \given X_1 = x_1] = \EX[U_Y \given U_1 = x_1].
\end{equation*}
The covariance $\sigma_{Y1}$ can then be obtained from $\EX[U_Y \given U_1]$ and $P(U_1)$.
\end{proof}

\subsection{MODEL ESTIMATION}
In this section, we propose a joint modeling approach to learn the parameters of the additive noise causal models from a combination of observational and interventional data. In particular, we define a likelihood for each data point according to its corresponding regime and then maximize the combined likelihood in which the model parameters are shared across different regimes and jointly optimized.

To define the combined likelihood, we first rewrite \eqref{eq:3.3} using vector notation where we define $X_{Y} := Y$ for notational convenience:
\begin{equation*}
    \bs{X} = \bs{f}(\bs{X}) + \bs{U}
\end{equation*}
with $\bs{X} = \mtx{
X_1 \\
\vdots \\
X_{Y}}
$, $\bs{f}(\bs{X}) = \mtx{
f_1(\textbf{PA}_{X_1}) \\
\vdots \\
f_{Y}(\textbf{PA}_{Y})
}$, $\bs{U} = \mtx{
U_1 \\
\vdots \\
U_{Y}}$.

By the change of variables, the distribution of the endogenous variables can be expressed in terms of the distribution of the noise variables as,
\begin{equation*}
    p_{\bs{X}}(\bs{x}) = p_{\bs{U}}(\bs{x} - \bs{f}(\bs{x}))\abs{\det(\bs{I} - \Delta \bs{f}(\bs{x}))}
\end{equation*}
where $\bs{I}$ denotes the identity mapping and $\Delta \bs{f}(\bs{x})$ denotes a Jacobian of $\bs{f}$ evaluated at $\bs{x}$. Since we assume that the underlying SCM is acyclic, the term $\abs{\det(\bs{I} - \Delta \bs{f}(\bs{x}))}$ is equal to 1. We can then define the likelihood of each data point according to its corresponding regime as follows.

\subsubsection*{Likelihood}

We are given an i.i.d sample of size $N_0$ from the observational regime denoted by $D_{obs} = \{\accentset{0}{\bs{x}}^{(n)}\}_{n=1}^{N_0}$ and a set of $K$ i.i.d samples of size $\{N_1, \dots, N_K\}$. Each of which is generated from the interventional regime where the treatment $X_k$ is intervened on, denoted by $\bs{D}_{int} = \{D^k_{int} = \{\accentset{k}{\bs{x}}^{(n)}\}_{n=1}^{N_k}\}_{k=1}^K$. In addition, we parameterize  functions $\bs{f}$ with $\bs{\theta}$, denoted by $\bs{f}(\cdot; \bs{\theta})$.

The likelihood for each data point in $D_{obs}$ is defined as,
\begin{align*}
    L^0(\bs{\theta}, \Sigma^0; \accentset{0}{\bs{x}}^{(n)}) = p_{\bs{U}}^0(\accentset{0}{\bs{x}}^{(n)} - \bs{f}(\accentset{0}{\bs{x}}^{(n)}; \bs{\theta}); \Sigma^0)
\end{align*}
where $p_{\bs{U}}^0 \sim \mc{N}(\bs{0}, \Sigma^0)$, and $\Sigma^0$ is a covariance matrix of size $(K + 1) \times (K + 1)$.

As for the interventional regime, the intervened variable $X_k$ is replaced by a constant, and thus the likelihood reduces to the joint probability of all other variables $\bs{X} \backslash X_k$ given the parameters. 

Define $\accentset{k}{\bs{x}}_{-k}^{(n)} := \accentset{k}{\bs{x}}^{(n)}\backslash \accentset{k}{x}^{(n)}_k, \bs{U}_{-k} := \bs{U} \backslash U_{k}, \bs{f}(\cdot)_{-k} := \bs{f}(\cdot) \backslash f_k(\cdot)$. The likelihood for each data point in $D^k_{int}$ can be defined as,
\begin{equation*}
    L^k(\bs{\theta}, \Sigma^k; \accentset{k}{\bs{x}}^{(n)}) = p_{\bs{U}_{-k}}^k(\accentset{k}{\bs{x}}_{-k}^{(n)} - \bs{f}_{-k}(\accentset{k}{\bs{x}}^{(n)}; \bs{\theta}); \Sigma^k)
\end{equation*}
where $p^{k}_{\bs{U}_{-k}} = \mathcal{N}(\bs{0}, \Sigma^{k})$, and $\Sigma^{k}$ is a covariance matrix of size $K \times K$.

We can then define the combined log-likelihood as, 
\begin{equation}
    \bs{\ell}(\bs{\theta}, \bs{\Sigma};D_{obs}, \bs{D}_{int}) = \sum_{k=0}^K \sum_{n=1}^{N_k} \log L^k(\bs{\theta}, \Sigma^k; \accentset{k}{\bs{x}}^{(n)}) \label{eq:com_llh}
\end{equation}
The maximum likelihood estimators for $\bs{\theta}$ and $\bs{\Sigma}$ can be obtained by maximizing the combined log-likelihood \footnote{We note that there could be a potential issue when the number of samples in the observational regime grows much faster than the interventional regimes, the observational regime could dominate the likelihood landscape and create undesired behavior during optimization. Our method is currently lacking treatment of such issues and we leave this for future work.}. To do so, we can alternately solve for $\bs{\theta}$ and $\bs{\Sigma}$ as follows.

\textbf{Solving for $\bs{\theta}$:}
Fixing the covariance matrices $\{\bs{\Sigma}\}$, the combined log-likelihood \eqref{eq:com_llh} can be maximized by standard iterative methods. The choice of an optimization algorithm can be chosen based on the choice of the functions $\bs{f}$. For instance, if we represent $\bs{f}$ by deep neural networks, we may consider commonly used gradient-based optimization algorithms such as Adam \citep{adam} or Adagrad \citep{adagrad}.

\textbf{Solving for $\bs{\Sigma}$:}
Fixing the functions $\bs{f}$, all the $K+1$ terms of the outer summation in \eqref{eq:com_llh} are disjoint and can thus be optimized separately. For each $k$, the problem boils down to an estimation of a multivariate normal covariance matrix for which the closed-form solution can be obtained (the maximum likelihood estimate of the covariance $\Sigma^k$ is essentially the sample covariance matrix).

\section{EXPERIMENTS}

\label{sec:exp}
In this section, we conduct multiple experiments to empirically illustrate the identifiability result, and the dangers of not properly modeling the likelihood function as a combination of different regimes. Furthermore, we compare the performance of our approach against a baseline approach on both synthetic and real-world datasets.

\subsection{BASELINE}

Given observational data $D_{obs}$ and sets of single-variable interventional data $\bs{D}_{int}$, a straightforward approach for estimating the effect of joint interventions is to pool all the data together $D_{pool} = \{D_{obs}, \bs{D}_{int}\}$ and directly fit a regression model on $D_{pool}$ to estimate $\EX[Y \given \textbf{PA}_Y]$. We denote this approach as $\texttt{REG}$ which is considered as our baseline throughout the experiments \footnote{We did consider another baseline where the regime indicators are used in the regression model. The results are consistently worse than $\texttt{REG}$ approach; therefore we decided not to include the results in the main paper. Please see the supplementary material for further details.}. Even though the baseline and our approach (denoted by \texttt{ANM}) utilize the same datasets, the main difference is that the baseline does not take into account the full causal structure (it partially uses the causal structure by considering only the direct causes of the outcome in the regression model), whereas our approach does exploit the full causal structure which is crucial in obtaining consistent estimates of the joint effects as we will see in the experimental results.

\subsection{SYNTHETIC EXPERIMENT}
\label{sec:syn-exp}
In this subsection, we illustrate our identifiability result on simulated data. Observational and interventional data are simulated according to a pre-defined SCM with correlated exogenous variables. We then compare the parameter estimates from our approach and the baseline on those simulated data. In particular, we empirically illustrate that our estimator is consistent. By using non-regularized polynomials in our function class, experiments will also empirically illustrate unbiasedness properties of our estimator even under a nonconvex optimization formulation. 

\subsubsection{Data Generating Process} In the synthetic experiment, we consider the case where the number of treatments $K = 3$. The data generating process is an additive noise SCM,

{\centering
  $ \displaystyle
\begin{aligned}
    Y &= f_Y(X_1, X_2, X_3; \bs{\theta}_y) + U_Y \\
    X_3 &= f_3(X_1, X_2; \bs{\theta}_3) + U_3 \\
    X_2 &= f_2(X_1; \bs{\theta}_2) + U_2 \\
    X_1 &= U_1
\end{aligned}
  $ 
\par}

where $(U_1, U_2, U_3, U_Y) \sim \mc{N}(\bs{0}, \Sigma_u)$. 

We examine both linear and non-linear structural equations $\bs{f}$ in our experiments. Specifically, the nonlinear functions considered are polynomials with second-order interactions in addition to the main effects. The same functional forms are also used in the baseline when fitting the direct regression models.

\subsubsection{Assessing Consistency and Unbiasedness} 
\begin{figure}[!t]
    \centering
    \subfigure
    {
        \includegraphics[scale=.102]{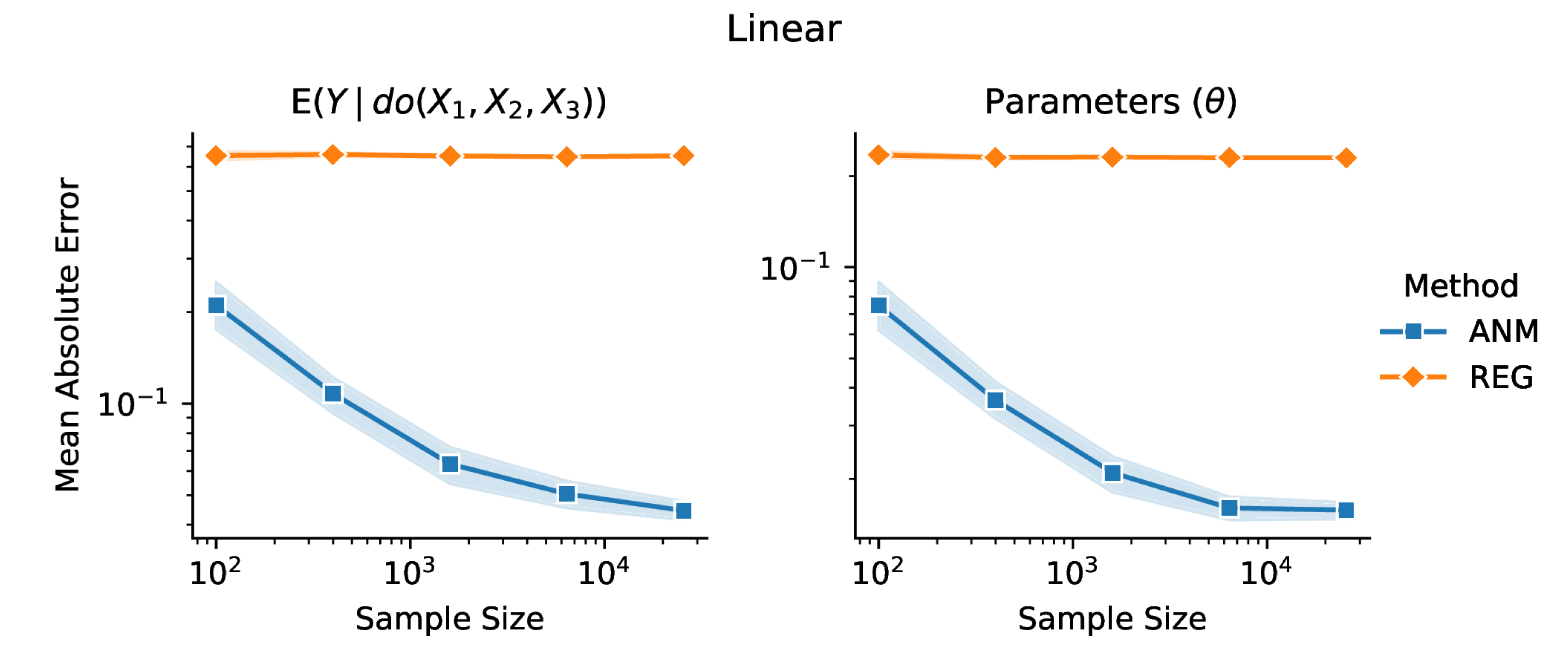}
    }
    \subfigure
    {
        \includegraphics[scale=.102]{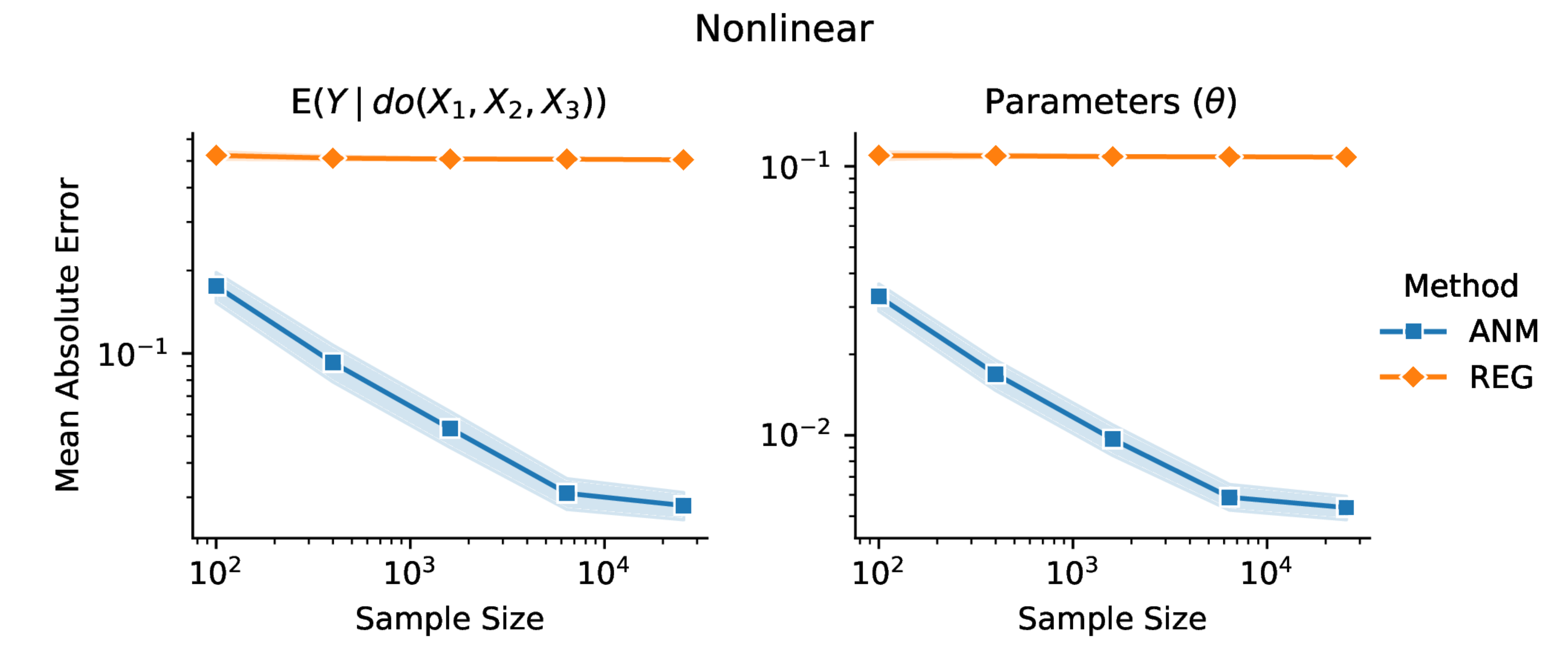}
    }
    \caption{MAE of the predicted joint effect (left) and the parameter estimates (right) as the sample size increases. The solid lines represent the mean absolute error averaged over 50 Monte Carlo experiments and the filled regions depict its 95\% confidence interval (note that both vertical and horizontal axes are in logarithmic scale).}
    \label{exp_fig:cons}
\end{figure}

We fix a simulated SCM with pre-defined parameters and evaluate consistency of our estimator by comparing the fitted values and the ground truth values of both parameter estimates and the predicted joint effects as the sample size increases ($n_{sample} = 100, 400, 1600, 6400, 25600$). For parameter estimates, we compute the absolute differences and average among all the parameters. For the predicted joint effects, we evaluate the mean absolute error (MAE) on 100000 uniform random joint intervention test points. We perform 50 simulations for each sample size. In each simulation, the observational data of size $n_{sample}$ is generated from the specified SCM. Furthermore, the intervention points $do(X_k = x_k)$ are randomly chosen according to the marginal distributions $P(X_k)$. The sets of $K$ \emph{single-variable} interventional data are then generated from the intervened SCMs, each of which has the size of $n_{sample}$. Figure \ref{exp_fig:cons} depicts the consistency result. For \texttt{ANM}, the errors both in terms of the predicted joint effect and the parameter estimates decrease and approach to zero as the sample size increase which indicates the consistency of our estimator. On the other hand, \texttt{REG} clearly suffers from inconsistency.

\begin{figure}[t!]
    \centering
    \subfigure
    {
        \includegraphics[scale=.095]{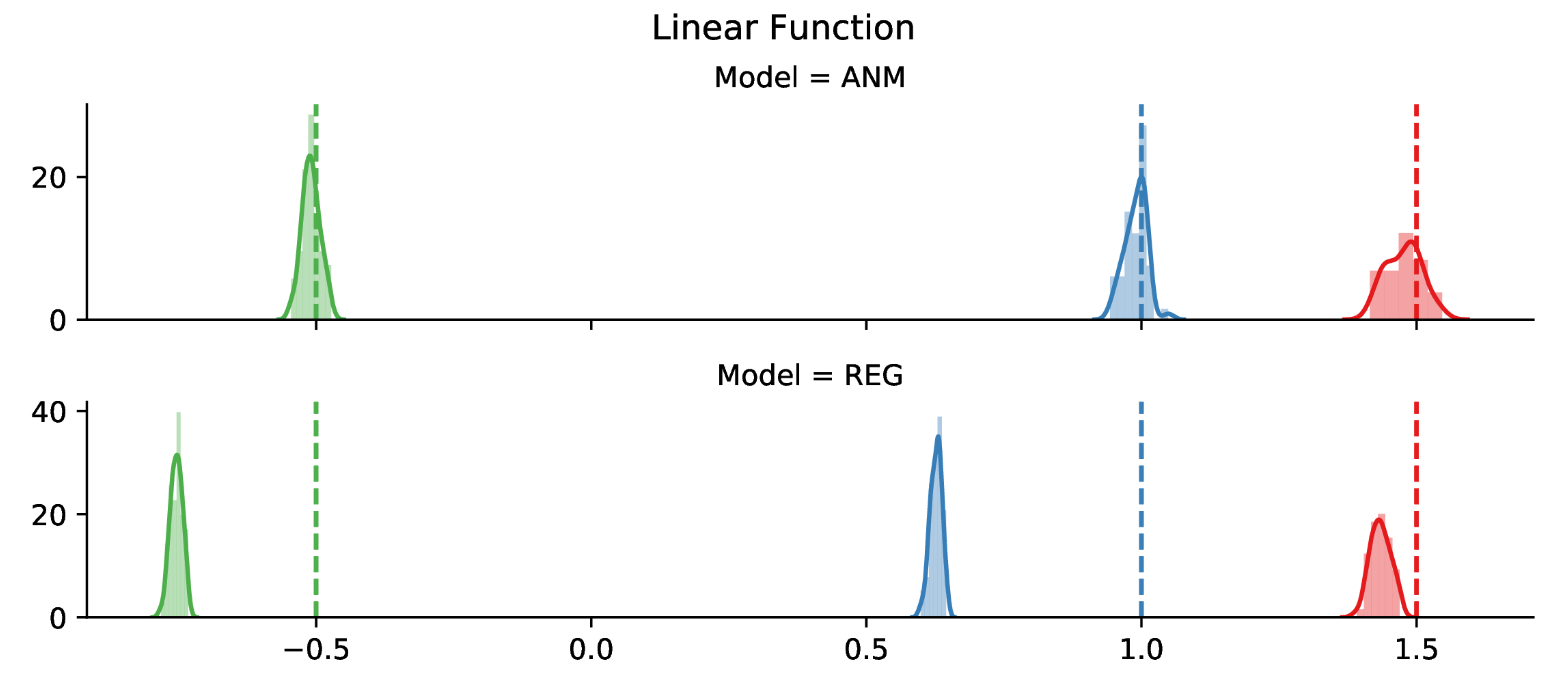}
    }
    \subfigure
    {
        \includegraphics[scale=.095]{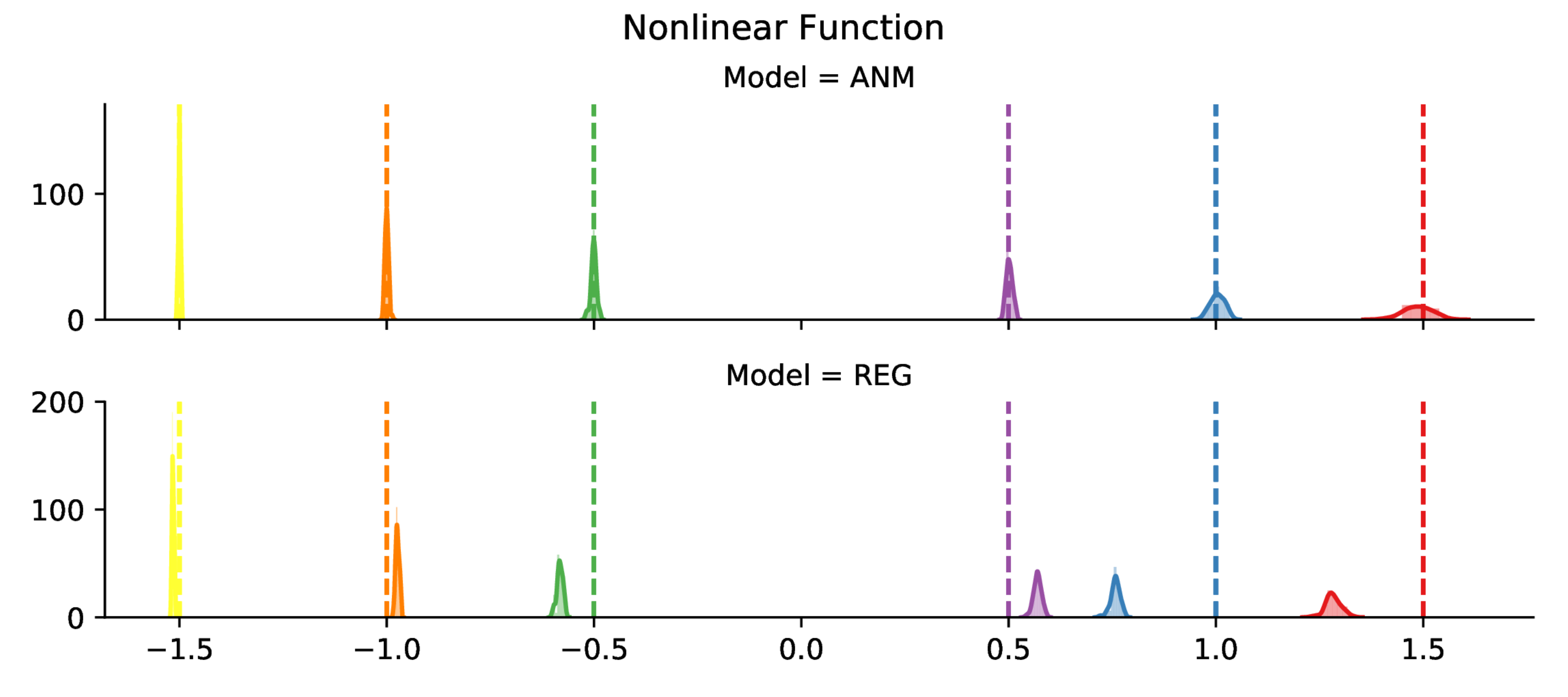}
    }
    \caption{Sampling distributions of the parameter estimates. Each parameter is coloured differently. The dot lines represent the ground truth parameter values.}
    \label{exp_fig:unbias}
\end{figure}

In addition to the consistency, we also empirically illustrate that our estimator can be unbiased\footnote{This will not happen in general if regularization is used.} by comparing the sampling distributions of the parameter estimates with the ground truth values for a fixed a sample size ($n_{sample} = 1600)$. The sampling distributions are simply obtained by Monte Carlo simulation. Figure \ref{exp_fig:unbias} shows the sampling distributions of the parameter estimates. In short, we can see that our model (\texttt{ANM}) produces unbiased estimates of the model parameters, whereas the parameter estimates obtained from the baseline (\texttt{REG}) are considerably biased regardless of sample size.

\subsection{SEMI-SYNTHETIC EXPERIMENT}
We further evaluate our proposed approach on data motivated by a real-world generative process. 
In order to have access to ground-truth, we create a semi-synthetic dataset by constructing a realistic data generating process (i.e. an SCM) from a dataset designed using expert knowledge, and use the learned SCM to generate plausible training and test data according to a best fit to an additive Gaussian error model. 


\subsubsection{Dataset and Setting}
We base our semi-synthetic experiment on a gene expression dataset from the DREAM4 challenge \citep{DREAM41,DREAM42,DREAM43}. The DREAM4 dataset consists of several gene regulation networks each of which are accompanied by several types of perturbation data, along with the gold-standard causal structure. In our experiment, we adopt one of the networks with 10 variables along with its corresponding multifactorial perturbations. The multifactorial perturbations are steady-state levels obtained after applying multifactorial perturbations which can be seen as gene expression profiles from different patients, i.e. observational data. 

The multifactorial perturbations data is then used to learn an underlying SCM based on the gold standard causal structure. The underlying SCM is an additive noise model with the nonlinear functional form described in the previous experiment. All the variables are standardized by subtracting the mean and dividing by the standard deviation. To simulate a confounding effect, we put a constraint on the correlation between the noise variables where the correlation matrix is randomly generated. Furthermore, as the gold standard causal structure may have cycles, we employ a heuristic algorithm proposed by \citet{EAEA93} to search for a minimal feedback arc set and remove them from the graph to obtain a DAG. Having obtained the underlying SCM, we then generate training data from the learned SCM where we define the outcome variable as the last node according to the topological ordering and all other nodes as treatment variables (the number of treatment variables $K=9$). Observational data and single-variable interventions are generated in the same way as described in Section \ref{sec:syn-exp}.

\subsubsection{Comparing the performance of the models}
\begin{figure*}[t!]
    \centering
    \subfigure[Varying Sample Size]{
        \includegraphics[scale=0.117]{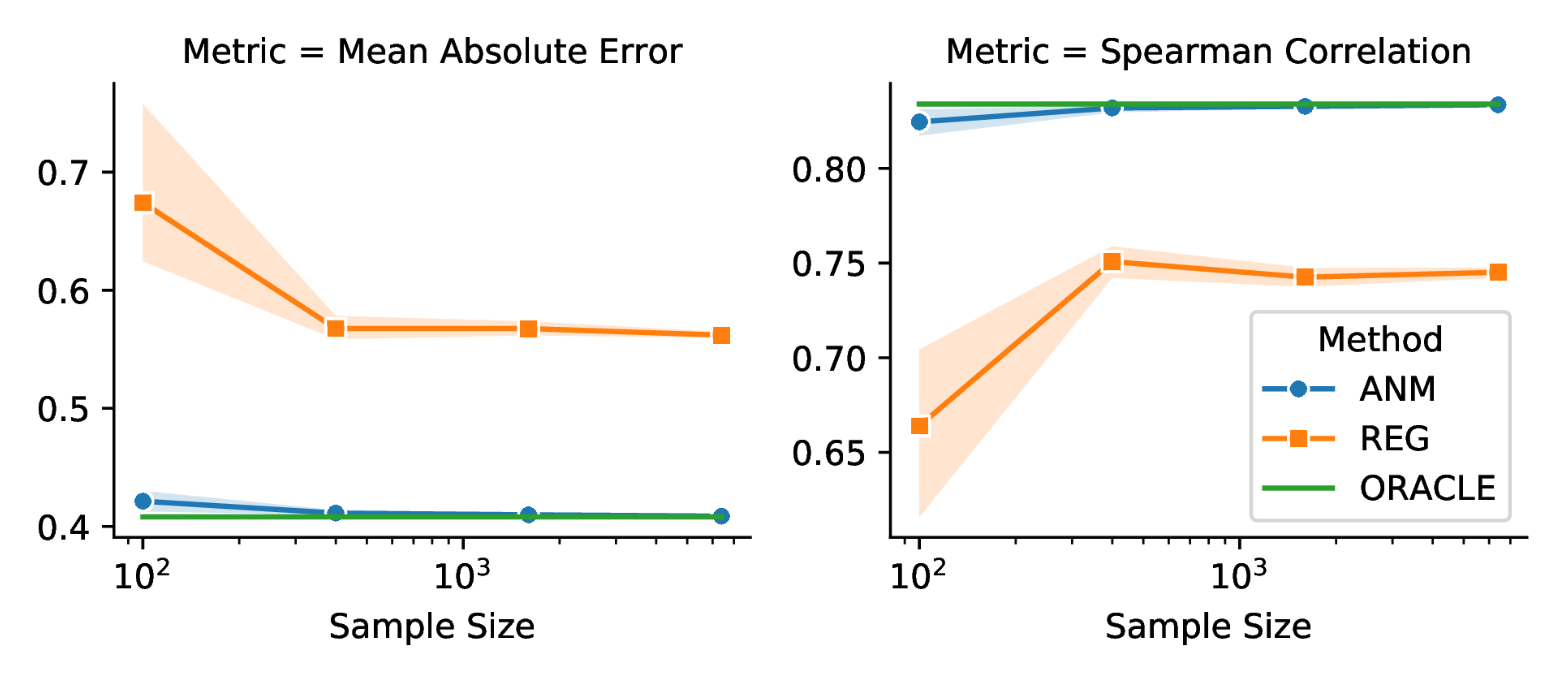}
        \label{exp_fig:real_acc}
    }
    \subfigure[Varying Confounding Level]{
        \includegraphics[scale=0.117]{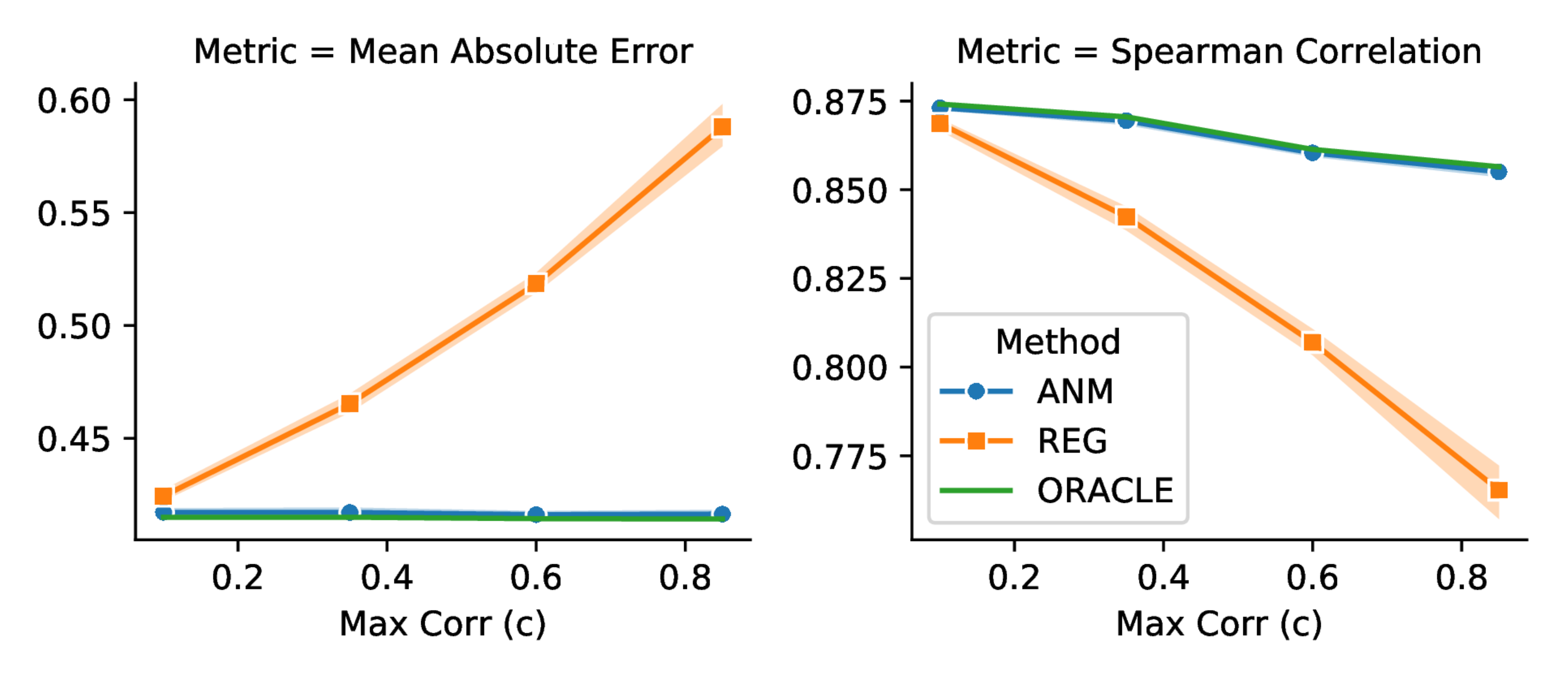}
        \label{exp_fig:real_conf}
    }
    \caption{MAE and Spearman's Correlation (a) as the sample size increases (b) as the confounding level increases. The solid lines represent the metric values, and the filled regions depict its 95\% confidence interval obtained from 50 bootstrap samples. The straight green lines represent the best possible values obtained when the ground-truth parameters are used in the model.}
\end{figure*}

We compare the performance of our model against the baseline as we vary the size of observational and single-variable interventional data ($n_{sample} = 100, 400, 1600, 6400$). The model performance is measured by how well they predict the effect of joint interventions on various test points. We generate a large number of uniform random joint intervention test points ($n_{test} = 100000)$ and compute two evaluation metrics, MAE and Spearman's Rank correlation between the predicted joint interventional effects and the ground truth values. We consider Spearman's Rank correlation in addition to MAE because, in many applications, we are interested in the ranking of the predicted interventional effects rather the actual predicted values. Figure \ref{exp_fig:real_acc} presents the results. Our approach (\texttt{ANM}) clearly outperforms the baseline (\texttt{REG}) in both metrics. Moreover, for a large enough sample size, the performance of \texttt{ANM} is almost on a par with the oracle's performance.

\subsubsection{Assessing the effect of hidden confounders}
We now investigate the impact of hidden confounders on our model and the baseline. We fix the number of samples $n_{sample} = 1600$ and vary the confounding levels of the underlying SCM by varying the magnitude of the correlation coefficients of the noise variables. Specifically, we limit the size of the non-diagonal entries in the correlation matrix by a constant $c$ (i.e. $\abs{\text{corr}(U_i, U_j)} \leq c$ for all $i \neq j$). We consider four confounding levels $c = 0.1, 0.35, 0.65, 0.8$. Figure \ref{exp_fig:real_conf} depicts the impact of the unobserved confounding on both models. Observe that when the confounding level is low, the performance of \texttt{ANM} and \texttt{REG} are relatively the same as expected. However, the performance of \texttt{REG} deteriorates drastically as we increase the confounding levels, whereas the one of \texttt{ANM} remains unaffected by the change in the confounding levels. This demonstrates that our approach is able to disentangle the confounding effects.

\begin{figure}[!b]
    \centering
    \includegraphics[scale=0.12]{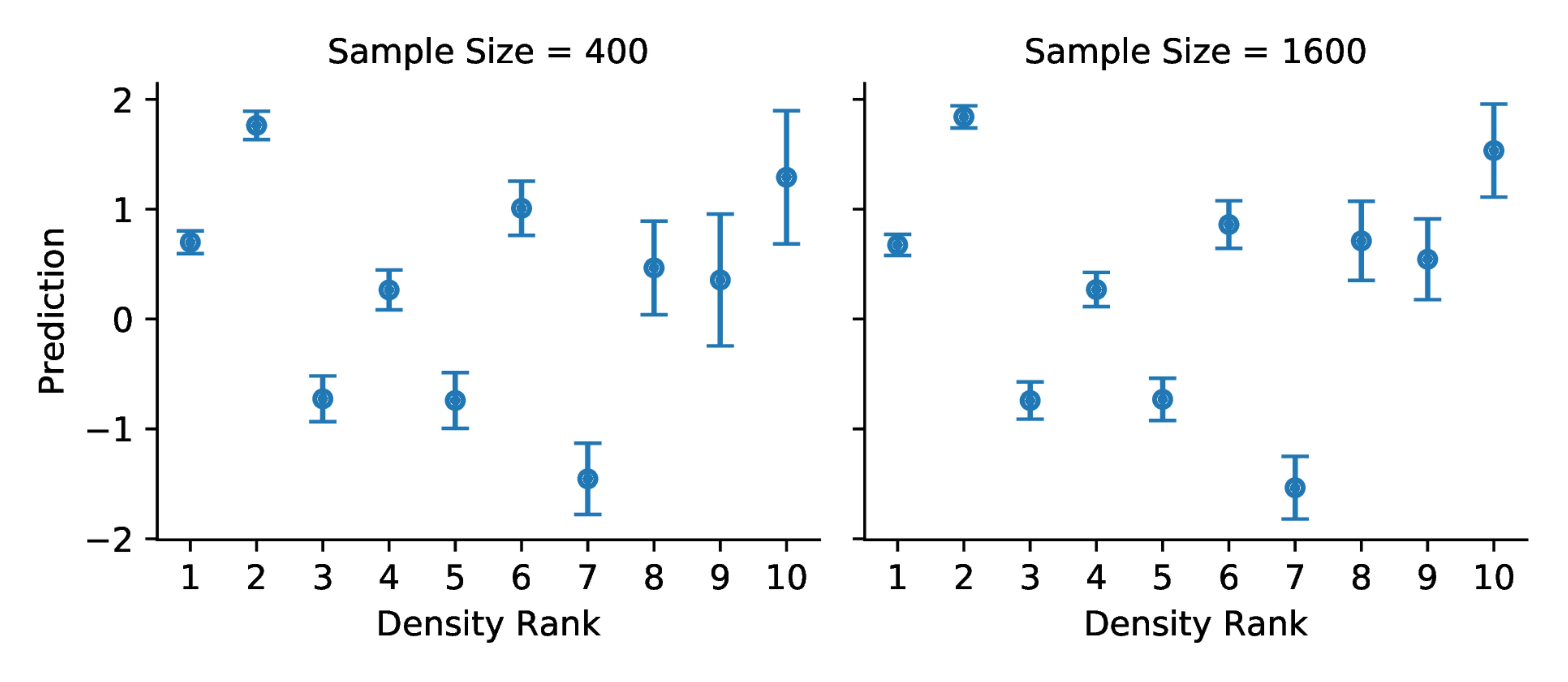}
    \caption{Prediction intervals of the \texttt{ANM} model. Each data point represents a test point where the x-axis represents how likely that test point is observed under the observational distribution (1:very likely, 10:very unlikely).}
    \label{exp_fig:real_uncertain}
\end{figure}

\subsubsection{Examining the uncertainty of the causal predictions}
In the last experiment, we examine the prediction uncertainty of our proposed model. We first select 10 uniform random joint intervention test points and rank them according to the probability of being generated from the observational distribution. In particular, we estimate the probability density function of the observed variables using kernel density estimation with a Gaussian kernel. The test points are then evaluated on the estimated density and get ranked accordingly. For each test point, we estimate prediction intervals of the joint effects based on 50 bootstrap samples. Figure \ref{exp_fig:real_uncertain} shows the prediction intervals on the selected test points. As expected, we have high uncertainty in our predictions when the test points fall within a low-density region of the distribution of the training data. This illustrates a potential issue when there is little overlap between intervention points in the training and the test data and that we need to be careful when making predictions in the low overlap regions.

\section{DISCUSSION}
\label{sec:diss}
In this paper, we proposed an approach to combine single-variable interventions for learning the effect of joint interventions. Our approach relies on relatively weak assumptions compared to existing methods and, most importantly, does not assume the absence of hidden confounders.

In future work, we plan to tackle the setting where we may observe context variables (covariates) in addition to the treatment variables. In this setting, we are often interested in learning interventional effects conditioned on the covariates which can be high-dimensional. We hypothesize that methods for estimating heterogeneous treatment effects 
\citep{NIEA17, CHEA18} 
can be integrated with our approach. 

Furthermore, we believe that this line of work could play significant roles in improving exploration in reinforcement learning/bandits. Our work provides a principled way to predict unseen interventions by exploiting the causal models, which in turn reduces the number of interventions needed in exploring the action space. How to effectively incorporate our approach in reinforcement learning/bandits is a promising future research direction.

\section{ACKNOWLEDGEMENTS}
Ricardo Silva has been supported by The Alan Turing Institute under EPSRC grant EP/N510129/1. The authors would also like to thank François-Xavier Aubet for his helpful feedback during the writing process.

\raggedbottom

\bibliographystyle{icml2020}
\bibliography{refs}

\end{document}